\documentclass[prb,twocolumn,amsmath,amssymb,showpacs]{revtex4}
\usepackage{graphicx}

\begin{document}

\title{Enhancement of pairing in a boson-fermion model for coupled
ladders}

\author{J. A. Riera} 
\affiliation{
Instituto de F\'{\i}sica Rosario, Consejo Nacional de
Investigaciones
Cient\'{\i}ficas y T\'ecnicas, y Departamento de F\'{\i}sica,\\
Universidad Nacional de Rosario, Avenida Pellegrini 250,
2000-Rosario, Argentina}

\date{\today}

\begin{abstract}
Motivated by the presence of various charge inhomogeneities in
strongly correlated systems of coupled ladders, a model of spatially
separated bosonic and fermionic degrees of freedom is numerically
studied. In this model, bosonic chains are connected to fermionic
chains by two types of generalized Andreev couplings. It is shown
that for both types of couplings the long-distance pairing
correlations are enhanced. Near quarter filling, this effect is much
larger for the
splitting of a pair in electrons which go to the two neighboring
fermionic chains than for a pair hopping process. It is argued that
the pairing enhancement is a result of the nearest neighbor Coulomb
repulsion which tunes the competition between pairing and charge
ordering.
\end{abstract}


\maketitle

\section{Introduction}
\label{intro}

Charge inhomogeneities are an ubiquitous feature in strongly correlated
electron systems. One of these inhomogeneities is the stripe phase
present in some underdoped cuprates.\cite{tranquada} In this phase,
it has been recently suggested by theoretical\cite{arrigoni} and
experimental\cite{tranq04} studies that both stripes and
intervening spin regions may be modeled as two-leg ladders.
Charge inhomogeneity can also be originated by the structure of the
materials. In the layered
compound $\beta$-Na$_{0.33}$V$_2$O$_5$, the V-O planes consist of
two-leg ladders separated by zig-zag chains.\cite{yamada} This
compound undergoes a transition from a charge-ordered 
state\cite{yamada,itoh} to a superconducting state under 
pressure.\cite{yamauchi} Within a purely electronic mechanism for
superconductivity in this material it is tempting to associate the
formation of pairs to the ladder units.\cite{drs} Similarly, in 
the compound Sr$_{14-x}$Ca$_x$Cu$_{24}$O$_{41}$, the Cu-O planes
consist of coupled ladders forming a trellis lattice. This material
becomes also superconductor under pressure\cite{uehara} and 
it also presents a competing charge ordered phase.\cite{fujiwara}

Then, it seems natural to study these inhomogeneous systems 
with Hubbard or $t$-$J$ Hamiltonians defined on coupled ladders or 
quasi-one dimensional structures, with in general different fillings.
After all, single ladders contains pairing, pseudogap, charge
ordering (CDW).\cite{drs,vojta} However, these models are quite
difficult
to analyze, both by analytical or numerical techniques. To make this
problem more manageable, effective models of bosons and fermions
can be derived and studied. These models may be obtained after a
basis change from the site to the dimer basis
\cite{sachdev,riera} or to the plaquette basis \cite{altman}
and eventually projecting out some states of the new basis.

Simplified models where the hard-core bosons describing triplet
excitations have been eliminated have also been studied 
recently.\cite{notriplet} These models were originally proposed
to describe bipolaronic superconductivity\cite{robasz} and later
used to study the pseudogap phase in the cuprates\cite{geshkenbein} 
where typically bosons represent preformed pairs.
In the present work, as a difference with those earlier studies,
a model in which bosonic and fermionic degrees of freedom
are {\em spatially separated} is considered.\cite{lehur}

In the case of the stripe order, as suggested by one of the main
theories about this phase,\cite{emerykiv} preformed pairs on 
the stripes would be described by hard-core bosons while unpaired
charge in the intervening regions would be described by fermions. 
Although pairing on stripes is a controversial 
issue,\cite{rierastrip} it is still interesting
to examine a possible enhancement of superconductivity due to
proximity effect within a simple lattice model.
In the same way, the compounds
with structural ladders could also be modeled by spatially separated
bosons and fermions. Bosons would describe pairs on ladders and 
fermions unpaired electrons on the zig-zag chains. In both cases,
a relatively simple boson-fermion model could give qualitative
features about, for example, the competition between 
superconductivity and CDW, the effect of strong electron correlations,
and the effect of 
applied pressure which leads to the modification of couplings, 
site energies.\cite{aronson}

In general, these effective models will contain fermion and boson
hopping terms together with some additional terms mixing bosons and
fermions. One of the most important and interesting mixing term is
a generalized Andreev coupling which describes the breaking of a
pair in two electrons or the reverse process. In Section \ref{model}
the model studied in this paper, formulated on a system of 
alternating bosonic and fermionic chains, is derived from microscopic
models using a projection technique. Results obtained by exact
diagonalization are shown in Section \ref{results}. Finally, the relevance
of these results to various physical systems is discussed in Section
\ref{conclusions}. 

\section{Effective model}
\label{model}

\begin{figure}
\begin{center}
\setlength{\unitlength}{1cm}
\includegraphics[width=3.75cm,angle=0]{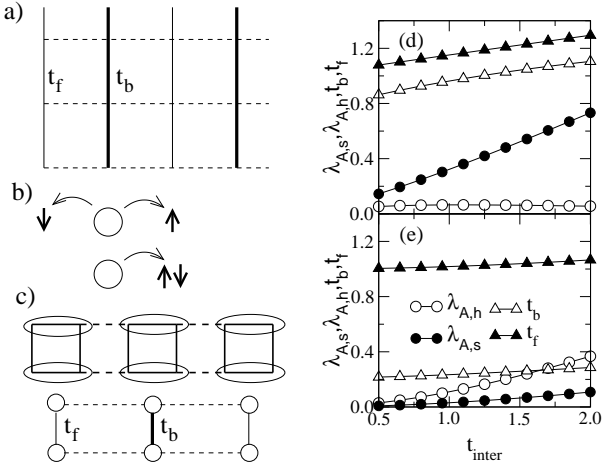}
\hspace{0.25cm}
\includegraphics[width=3.75cm,angle=0]{./fig1de.eps}
\caption{(a) Inhomogeneous system of coupled chains with bosonic
(thick lines) or fermionic degrees of freedom (thin lines).
(b) Two different processes of creation/annihilation of pairs:
splitting (top) and pair hopping (bottom). (c) derivation of an
effective model by changing to a dimer basis followed by projection.
(d) effective couplings (defined in the text) as a function of the
inter-dimer hopping, $J=0.4$, $V_0=0$ and $t=1$ in the original 
12-site cluster. (e) Same as (d) for $U_0=-8$, $V_0=2$ and $t=1$.
}
\label{fig1}
\end{center}
\end{figure}

The specific model studied in this paper is formulated on a system of 
alternating bosonic and fermionic chains as shown in 
Fig.~\ref{fig1}(a). The Hamiltonian is defined as:
\begin{eqnarray}
{\cal H} = &-& t_{b} \sum_{<i,j>} (b^{\dagger}_{i} b_{j} + h.c. )
- t_f \sum_{<i,j>,\sigma }
(c^{\dagger}_{i \sigma} c_{j \sigma} + h.c. )  \nonumber \\
&+& U \sum_{i} n_{i \uparrow} n_{i \downarrow} +
V \sum_{<i,j>} e_{i} n_{i} e_{j} n_{j} \nonumber \\
&+& \sum_{i} \epsilon_i n_{i}
+ \lambda_A \sum_{i,j,k} ( b^{\dagger}_{i} c_{j \uparrow}
c_{k \downarrow} + h.c. )
\label{hambf}
\end{eqnarray}
\noindent
where, $c^{\dagger}_{j\sigma}$ creates an electron with spin $\sigma$
at site $j$, $n_{j\sigma}=c^{\dagger}_{j\sigma} c_{j\sigma}$;
$n_{j}=n_{j \uparrow} + n_{j \downarrow}$ for fermions or
$n_{j}=b^{\dagger}_{j} b_{j}$ for hard-core bosons. The charge is
$e_{i}=1~(2)$ for fermions (bosons). 
$t_b$ and $t_f$ are the hopping integrals
along the chains for bosons and fermions 
respectively; $U$ is the onsite Coulomb repulsion on the
fermion sites; $V$ is the Coulomb repulsion on nearest neighbor (NN)
sites, acting between paired and/or unpaired electrons;
$\epsilon_i$ is the onsite energy at site $i$ which we take
$\epsilon_i=\epsilon$ on the bosonic chains, $\epsilon_i=0$ on
the fermionic chains. Thus, $\epsilon$ contains the binding energy.

The last term of the boson-fermion Hamiltonian Eq.(\ref{hambf}) is
a generalized Andreev coupling, transversal to the chains,
where $i$ is a site on a bosonic chain and $j$, $k$ are the 
NN sites of $i$ on the fermionic chains. Two different 
processes will be studied (Fig.~\ref{fig1}(b)). The one at the
top corresponds to a splitting of a pair into two electrons 
located in the two fermionic neighboring chains ($\lambda_{A,s}$)
while the one at the bottom corresponds just to a pair hopping to 
one of the two neighboring fermionic chains 
($\lambda_{A,h}$).\cite{note1}

In order to understand
the physical origin of these two types of Andreev coupling let us
consider an extended $t$-$J$ or Hubbard model on the 12-site system
of Fig.~\ref{fig1}(c) top with 2 electrons.  By performing a site
to dimer change of basis, and projecting out the one-electron
(fermionic) dimer states on the two inner dimers and the
double occupied (bosonic) states 
on the four outer dimers, one gets the effective 6-site 
boson-fermion model of Fig.~\ref{fig1}(c) bottom. 
The effective Hamiltonian is given by the standard formula:
\begin{eqnarray}
{\cal H}_{eff} = {\cal P H_{\rm 0} P} -
{\cal P H_{\rm 0} Q}\frac{1}{{\cal Q H_{\rm 0} Q}-E_0}
{\cal Q H_{\rm 0} P}
\label{project}
\end{eqnarray}
where ${\cal P}$ is the projection operator on the subspace of 
retained states, ${\cal Q }$ is the projection operator on the
subspace of the eliminated states, ${\cal H_{\rm 0}}$ is the
original Hamiltonian, ${\cal H_{\rm 0}} \Psi_0 = E_0 \Psi_0$, and
${\cal H}_{eff} {\cal P} \Psi_0 = E_0 {\cal P} \Psi_0$.\cite{fulde}
Variants of this procedure were repeatedly 
performed\cite{sachdev,riera,altman} to obtain
effective models from the Hubbard or t-J models, specially
retaining triplet states and projecting out fermionic states.
Similar studies but retaining fermionic states, although using
a different projection procedure\cite{altman}, have concluded that
the most important interactions not involving triplets are
the ones contained in Hamiltonian (\ref{hambf}). Although 
Eq. (\ref{project}) is usually analytically calculated using second
order perturbation theory, for the 12-site cluster of 
Fig.~\ref{fig1}(c) it could easily be numerically solved using 
standard matrix inversion subroutines.\cite{code} In the mapping
shown in Fig.~\ref{fig1}(c), the projection step also leads to
second neighbor fermion-fermion and three-site interactions in
the horizontal direction which are also negligible in first
approximation.

Let us consider first that ${\cal H}_{\rm 0}$ is an extended $t$-$J$
model.  Let us assume that we have the usual $t$, $J$ couplings and
a NN Coulomb repulsion $V_0$ on the ladders, and $t_{inter}$,
$J_{inter}$, $V_{inter}$ between ladders, with
$J_{inter}/J=(t_{inter}/t)^2$ and $V_{inter}/V_0=t_{inter}/t$.
It is reasonable to assume that the main effect of applying pressure
is to modify the interladder interactions, which are here related to
$t_{inter}$. 
The values of the effective couplings $t_b$, $t_f$, $\lambda_{A,s}$
and $\lambda_{A,h}$ are shown in Fig.~\ref{fig1}(d) for $J/t=0.4$,
$V_0J/t=0$, as a function
of $t_{inter}/t$. The most important feature is that 
$\lambda_{A,s}$ is clearly larger than $\lambda_{A,hop}$. It
should be noticed that the sign of $\lambda_{A,s}$ corresponding
to the process in Fig.~\ref{fig1}(b) is opposite to the process
in which the up spin electron goes to the left chain and the 
down spin electron goes to the right chain. This is related to
the fact that the electron pair on a ladder rung form a singlet.
One should also notice that $t_b$ and $t_f$ take values close to 
each other.

On the
other hand, let us assume that we have an extended Hubbard model
with $U_0<0$, $V_0>0$, on the 12-site
cluster of Fig.~\ref{fig1}(c). This attractive $U_0$ could describe
a phonon mediated pairing. In this case, one obtains the effective
parameters shown in Fig.~\ref{fig1}(e). It is now apparent that
the pair hopping type of Andreev coupling dominates over the pair
splitting type.  It should also be noticed that in this
case $t_b$ is much smaller than $t_f$
and hence it could be neglected which is precisely what has been
done in the earlier literature\cite{robasz,notriplet}
although it was included in the studies related to the pseudogap
phase in cuprates.\cite{geshkenbein}
It is possible to think then that the case considered in 
Fig.\ref{fig1}(d), with a dominance of $\lambda_{A,s}$, 
corresponds to a strongly correlated electron physics with
a likely d-wave pairing, while the situation of 
Fig.\ref{fig1}(e), would correspond to a more conventional,
s-wave, type of superconductivity.

In order to support this interpretation, the probability of double
occupancy and the probability of having a singlet on a rung,
properly normalized (i.e., the sum of all the possible configurations
on a rung equal to 1) was computed. In the case of the extended $t-J$
model, the probability of electrons forming a rung singlet is much
larger than the probability of electrons going to double-occupied sites.
The reverse situation occurs for the attractive Hubbard model. It is
instructive to consider also a model which interpolates between the
Hubbard and the $t$-$J$ model, the so-called $t$-$J$-$U$ model,
obtained from the $t$-$J$ model by relaxing the no double-occupancy
constraint but including an onsite Hubbard repulsion. For an
intermediate situation, for example, $J=1$, $t=1$, $U=0.5$, $V=0$,
there is a crossover from the splitting to the pair-hopping types
of Andreev coupling as $t_{inter}$ is increased consistent with a
crossing from singlet to double-occupancy order. That is, applied
pressure can change a s-wave pairing into a d-wave pairing.

The proposed model (\ref{hambf}) is far more general than the simple
``derivation" schematically shown in Fig.~\ref{fig1}(c). In the first
place, a boson does not necessarily represent a pair on a ladder rung.
In fact, it has been shown that on ladders, pairs are located along 
plaquette diagonals or on more distant sites depending on coupling
values.\cite{riera98} In general, this boson-fermion model is
applicable to any
compound containing quasi-1D units bearing some kind of pairing. The
site energy $\epsilon_i$ would be in general determined by the
binding energy of electrons on these quasi-1D units  as well as by a
potential coming from the whole structure of the material, which can
be modified by external applied pressure or by internal chemical
substitution, as in Sr$_{14-x}$Ca$_x$Cu$_{24}$O$_{41}$, where Ca
doping leads to transfer of holes from the chain to the ladder 
planes. The boson-fermion model could provide insights to
predict the effects produced by these kinds of perturbations on a 
given compound.

\section{Numerical results}
\label{results}

\subsection{$3\times L$ cluster, quarter filling}

Model Eq.(\ref{hambf}) was studied by exact diagonalization (Lanczos
algorithm) on
$3\times L$ ($L=6,8$) clusters with periodic (open) boundary
conditions (BC) along (across) the $L$-site chains. The central 
$L$-site chain has bosonic operators while the two external 
chains contain the fermionic ones. $t_f$ is adopted as the unit of
energy. In the above mentioned basis change, the effective 
parameters $U$ and $V$ result roughly half the NN Coulomb repulsion
of the original model ($V_0$), both for the limits of infinite and
zero values of the Hubbard on-site repulsion of the original 
model ($U_0$). Hence $U=V$
in the following. Various properties, specially those related
to superconductivity, were computed as a function of $t_b$,
$\epsilon$ and $\lambda_A$. 

\begin{figure}
\begin{center}
\setlength{\unitlength}{1cm}
\includegraphics[width=7.5cm,angle=0]{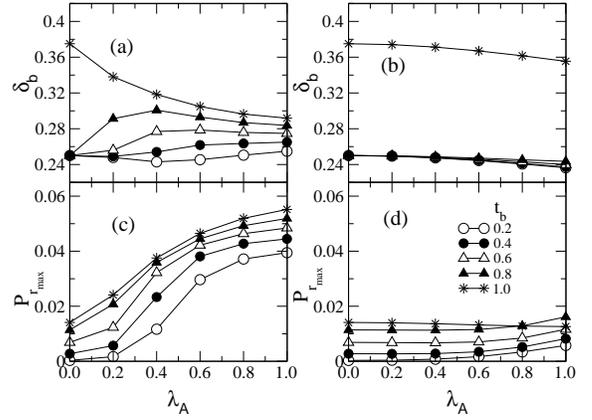}
\caption{Results for the $3\times 8$ cluster, $U=V=2$, 
$\epsilon=0$, $n=0.5$, as a function of $t_b$ and $\lambda_A$.
Boson density in the central chain: (a) pair splitting, (b) pair
hopping. Pair-pair correlations at the maximum distance along 
the central chain:  (c) pair splitting, (d) pair hopping.
}
\label{fig2}
\end{center}
\end{figure}

Figure \ref{fig2} shows results for the $3\times 8$ cluster,
$\epsilon=0$, $U=V=2$, at quarter
filling ($n=0.5$). The first feature to notice is that the relative
occupation of the fermionic and bosonic chains depends on the
parameters of the model. In Figs.~\ref{fig2}(a) and (b), the boson
density in the central chain, $\delta_b=\langle n_b \rangle /L$
($L=8$ in this case) is shown for $\lambda_{A,s}$
and $\lambda_{A,h}$ respectively. In both cases, $\delta_b$
increases as $t_b$ is increased. This is expected since electrons
would move to the central chain to gain kinetic energy.
For $t_b=1$ there is a level
crossing with a sudden increase of $\delta_b$.
On the other hand, $\delta_b$ slowly decreases with $\lambda_{A,h}$ 
in the interval shown, while its behavior with $\lambda_{A,s}$ 
is non monotonic.

The central quantity of the present study is the boson
correlation at the maximum distance, 
$P_{r_{max}}=\langle b^\dagger_{r_{max}} b_0\rangle$ 
along the central chain. This correlation, which in the current
model has the meaning of pairing correlation, is a measure
of quasi-long range superconducting order on the bosonic chain.
The results for $P_{r_{max}}$ are shown in Fig.~\ref{fig2}(c),(d)
for $\lambda_{A,s}$ and $\lambda_{A,h}$ respectively. In both 
cases, for a fixed value of $t_b$, $P_{r_{max}}$ shows an 
enhancement as the Andreev coupling increases. The curves of 
$P_{r_{max}}$ vs. $\lambda_A$ are shifted upward as $t_b$
increases, as expected. The most important feature of these
results is that the enhancement with $\lambda_A$ is much 
stronger for the case of pair splitting than for the case of 
pair hopping. Notice also that for pair splitting, for $t_b$
fixed, the behavior of $P_{r_{max}}$ and the one of $\delta_b$ are
unrelated. 
It should be emphasized that in the region where $P_{r_{max}}$ 
is enhanced, the pairing correlations as a function of distance,
$\langle b^\dagger_{r} b_0\rangle$, have a monotonically
decreasing behavior corresponding to true long-distance pairing.
A non-monotonic behavior would be indicative of phase separation
or CDW.

For $U=V=0$, energy, boson occupancy and pairing correlations are
identical for $\lambda_{A,s}$ and $\lambda_{A,h}$. This comes
from the fact that the respective Hamiltonians are related by
the transformation 
${\cal H}^{(h)}_0={\cal T}^{-1} {\cal H}^{(s)}_0{\cal T}$, 
where ${\cal T}$ relates the two processes depicted
in Fig.~\ref{fig1}(b) as 
${\cal H}_h={\cal T}^{-1} {\cal H}_s{\cal T}$, and
${\cal T}^{-1} ={\cal T}^{T}$. Quantities defined solely in terms 
of bosonic operators are preserved by this transformation. The 
important result is that in this case, $P_{r_{max}}$ is {\em not}
enhanced by $\lambda_A$, although it is considerably larger than 
for $U=V=2$.

\begin{figure}
\begin{center}
\setlength{\unitlength}{1cm}
\includegraphics[width=7.5cm,angle=0]{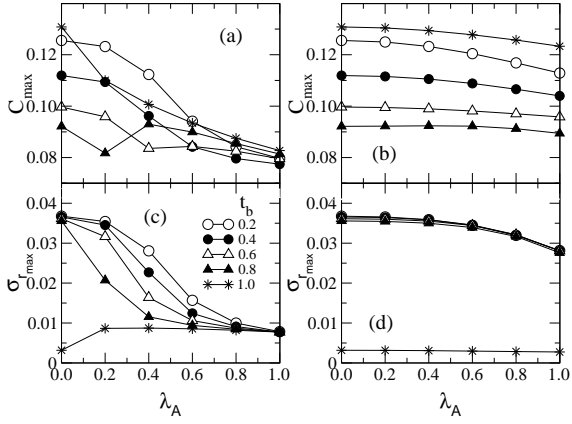}
\caption{Results for the $3\times 8$ cluster, $U=V=2$, 
$\epsilon=0$, $n=0.5$, as a function of $t_b$ and $\lambda_A$.
Maximum charge structure factor: (a) pair splitting, (b) pair
hopping. Current-current correlations at the maximum distance 
along the fermionic chains:  (c) pair splitting, (d) pair hopping.
}
\label{fig3}
\end{center}
\end{figure}

In order to characterize the physics of this model more completely,
the static charge structure factor $C({\bf q})$ and the 
current-current correlations at the maximum distance along one of 
the fermionic chains were computed. The first quantity is indicative
of CDW while the second one is related to the conduction of the 
fermionic chains.\cite{rierafut} For the same
parameters as before ($U=V=2$, $\epsilon=0$, $n=0.5$), $C({\bf q})$ 
presents a maximum at ${\bf q}_{max}=(3\pi/4,2\pi/3)$ in the whole
range of $t_b$ and $\lambda_A$ examined. Figures ~\ref{fig3}(a),(b) 
show $C_{max}=C({\bf q}_{max})$ for $\lambda_{A,s}$ and 
$\lambda_{A,h}$ respectively. It can be seen that this quantity is 
suppressed, particularly by $\lambda_{A,s}$. The charge
structure factor for the whole cluster behaves in a similar way
than the one computed from the charge-charge correlations along
a single fermionic chain. Now, for $U=V=0$, $C({\bf q})$, as it
was found for $P_{r_{max}}$, is roughly independent
of $\lambda_A$. The same behavior is also found for other clusters
and densities considered below. It is possible then to sum up these
features by suggesting that $\lambda_{A,s}$ works against the tendency
to CDW, favored by $V$, thus leading to an enhancement of
long distance pairing.

The current operator between sites $i$ and $i+\hat{y}$ is defined
as usual as $j_{\hat{y},i}= i e t \sum_{\sigma }
(c^{\dagger}_{i+\hat{y} \sigma} c_{i \sigma} - h.c. )$,
and current-current correlations at the
maximum distance as $\sigma_{r_{max}}=\langle j_{\hat{y},r_{max}} 
j_{\hat{y},0} \rangle$. Results for $\sigma_{r_{max}}$ 
along a fermionic chain are shown in Fig.~\ref{fig3}(c),(d) for 
$\lambda_{A,s}$ and $\lambda_{A,h}$ respectively. It can be seen
that $\sigma_{r_{max}}$ is suppressed in both cases although this
effect is much larger for $\lambda_{A,s}$ than for $\lambda_{A,h}$.
This behavior indicates that the effect of $\lambda_A$ is to
favour the conduction mainly through the bosonic chain.

\subsection{Other clusters and fillings}

It is also important to determine if the behavior shown in 
Figs.~\ref{fig2} and \ref{fig3} is also present
at larger electron fillings, specially because some possible
applications of the present model, for example
Sr$_{14-x}$Ca$_x$Cu$_{24}$O$_{41}$, correspond to 
systems close to half-filling. As electron filling
increases from $n=0.5$, the dimension of the Hilbert space increases
rapidly and it soon makes this problem very hard for exact
diagonalization. Hence we have to limit the study to the smaller
$3\times 6$ cluster.

\begin{figure}
\begin{center}
\setlength{\unitlength}{1cm}
\includegraphics[width=7.5cm,angle=0]{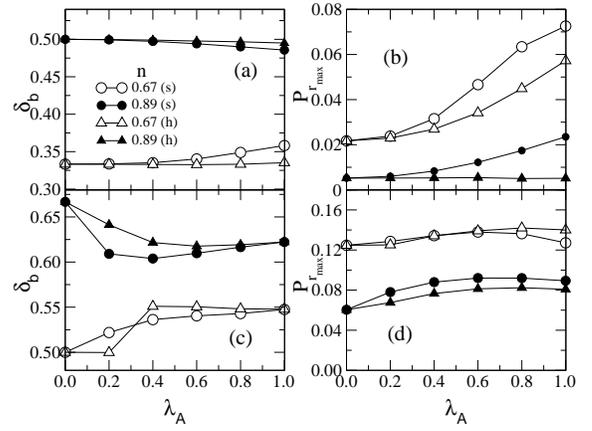}
\caption{Results for the $3\times 6$ cluster, $t_b=0.8$,
$n=0.67$ and 0.89.
(a) Boson density and (b) pair-pair correlations at
the maximum distance along the central chain  for
$U=V=1$, $\epsilon=-0.5$, as a
function of $\lambda_A$. Results for pair splitting 
(hopping) are indicated with circles (triangles). In (c) and (d),
the same quantities as in (a) and (b) respectively are shown for 
$U=V=0.5$, $\epsilon=-2$.
}
\label{fig4}
\end{center}
\end{figure}

Results for the $3\times 6$ cluster with 12 electrons ($n=0.67$)
and 16 electrons ($n=0.89$) are shown in Fig.~\ref{fig4}. For
$U=V=1$, $\epsilon=-0.5$, $t_b=0.8$, the boson density varies slowly
with $\lambda_A$ (Fig.~\ref{fig4}(a)). At $n=0.67$,
$\delta_b\sim 0.35$, which implies an identical charge density 
on fermionic and bosonic chains. At $n=0.89$, 
$\delta_b\sim 0.5$ implying a larger charge density on the bosonic
than on the fermionic chains. The overall behavior of the pairing
correlation at the maximum distance along the central chain,
shown in Fig.~\ref{fig4}(b), is the 
same as in Fig.~\ref{fig2}, i.e., there is an
enhancement of $P_{r_{max}}$ with $\lambda_A$ which is more 
pronounced for pair splitting process. For these values of the
parameters, $C_{max}$, peaked at ${\bf q}_{max}=(\pi,\pi)$ is
monotonically suppressed by $\lambda_A$.

To obtain larger charge density on the bosonic chain,
the couplings $U=V=0.5$ and $\epsilon=-2$ were studied; $t_b=0.8$ as
before. It can be seen in Fig.~\ref{fig4}(c), that at $n=0.67$,
$\delta_b$ becomes slightly larger than 0.5 and at $n=0.89$, 
$\delta_b\sim 0.6$. For both global fillings, the charge density
on the bosonic chain is approximately twice the one on the fermionic
chains. Notice that now $P_{r_{max}}$ (Fig.~\ref{fig4}(d)) is larger
than the one shown in Fig.~\ref{fig4}(b). This larger value, for
$n=0.89$, could be related to the behavior of simple hard-core boson
(or spinless fermion) chain with NN repulsion, where
superconductivity is suppressed at half-filling ($\delta_b=0.5$),
which is the case for $U=V=1$ and $\epsilon=-0.5$
(Fig.~\ref{fig4}(a)).
On the other hand, the opposite happens for the case of $n=0.67$,
indicating that the Andreev coupling changes the physics of an 
isolated bosonic chain. However, at $\lambda_A \leq 0.2$,
the pairing correlations as a function of distance has a 
non monotonic behavior, signaling CDW. It should also
be noticed that for all cases in Fig.~\ref{fig4}(d), 
there is a saturation and further decreasing of $P_{r_{max}}$ for
larger $\lambda_A$. This may indicate that the behavior shown in
Figs.~\ref{fig2}(c) and \ref{fig4}(b) is mostly a property of
{\em low} bosonic density ($\delta_b<0.5$).
It should be stressed that by going
from $\epsilon=-0.5$ to $-2$, with $U=V=1$ fixed, the changes are
smoother than by going from  $U=V=1$ to  $U=V=0.5$, keeping
$\epsilon=-2$ fixed. In this parameter space, results interpolate
smoothly between those of Fig.~\ref{fig4}(a),(b) and those of
Fig.~\ref{fig4}(c),(d).

\begin{figure}
\begin{center}
\setlength{\unitlength}{1cm}
\includegraphics[width=7.5cm,angle=0]{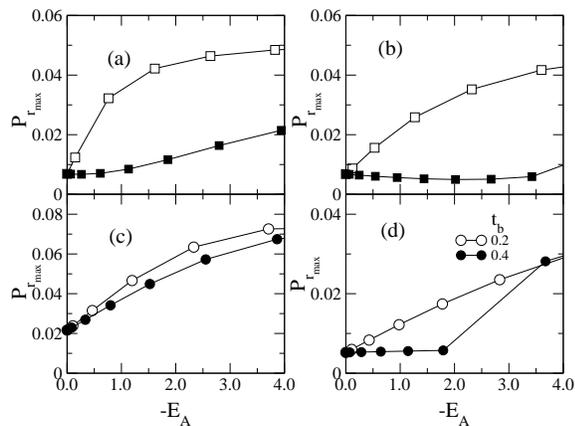}
\caption{Pair-pair correlations at the maximum distance along
the central chain versus the absolute value of the energy of the
Andreev term. Results for the $3\times 8$ cluster, $U=V=2$, 
$n=0.5$, $t_b=0.6$ (a) $\epsilon=0$, (b) $\epsilon=-0.5$.
Results for the $3\times 6$ cluster, $U=V=1$,  $t_b=0.8$,
$\epsilon=-0.5$, (c) $n=0.67$, (d) $n=0.89$.
}
\label{fignew}
\end{center}
\end{figure}

It is tempting to relate the relatively small effect of pair
hopping on pairing correlations to its possible suppression by
the onsite Coulomb repulsion on the fermionic chains. This is
actually not the case. One should notice first that pair 
splitting is also affected by such repulsion since an electron 
could be already present in one or both of the final sites on 
the fermionic chains. For noninteracting electrons simple 
combinatorics lead to the result that double-occupancy is more
likely on the splitting than on the pair processes for electron
densities larger than $\sim 0.6$. A similar effect caused by
the NN repulsion $V$ is more difficult to predict. Alternatively,
it is possible to compute the contribution of the Andreev term
to the total energy, $E_{A}$, as a measure of how much this term 
is actually ``working". It may be convenient then to plot 
$P_{r_{max}}$ as a function of $E_{A}$, rather than as a function of
the bare parameter $\lambda_{A}$. This is done in Fig.~\ref{fignew}
for some typical cases of Figs.~\ref{fig2} and \ref{fig4}. It
can be seen that the qualitative behavior of these figures is not
modified. Only in Fig.~\ref{fignew}(d), corresponding to a density
$x=0.89$,  there is a jump
in $P_{r_{max}}$ for the pair hopping case but this occurs at a
rather large value of the bare coupling, $\lambda_{A,h}=1.2$.
Figs.~\ref{fignew}(a) and (b) allows the comparison between two
different values of $\epsilon$. For both types of Andreev couplings,
a smaller (negative) value of $\epsilon$ gives a smaller enhancement
of $P_{r_{max}}$, an even a suppression in the case of the
pair type of Andreev coupling.

Computer limitations make it difficult to go to larger clusters
in order to assess finite size effects but it is possible to
study clusters with different geometry. The $2\times 12$ cluster
was considered to estimate finite size effects on the pair 
hopping between the fermionic and bosonic
chains. Results for the same parameters as in Fig.~\ref{fig2} 
show that $P_{max}$ takes values very close to those found for 
the $3\times 8$ cluster with a very similar (small) dependence
with $\lambda_{A,h}$. This asymmetric two-chain system is 
essentially the same studied by Le Hur\cite{lehur} by
bosonization techniques, although in this work the boson
mediated pairing of unpaired electrons is mainly investigated.
Finally, the cluster with 4 coupled chains of length 6, with
periodic BC also in the transversal direction, with 12 and 14
electrons, and the same couplings as in Figs.~\ref{fig2} and
\ref{fig3} was considered. The overall behavior is the same as
that depicted in those Figures -in particular $P_{r_{max}}$ is
much more robust for $\lambda_{A,s}$ than for $\lambda_{A,h}$-
with the additional feature of an enhancement of pairing
correlations also in the direction perpendicular to the
chains.\cite{rierafut} For $U=V=0$, as for the $3\times L$
clusters studied above, $C({\bf q})$ and longitudinal
$P_{r_{max}}$ are almost independent of $\lambda_A$
although transversal pairing correlations are trivially
enhanced by $\lambda_A$.

\begin{figure}
\begin{center}
\setlength{\unitlength}{1cm}
\includegraphics[width=7.5cm,angle=0]{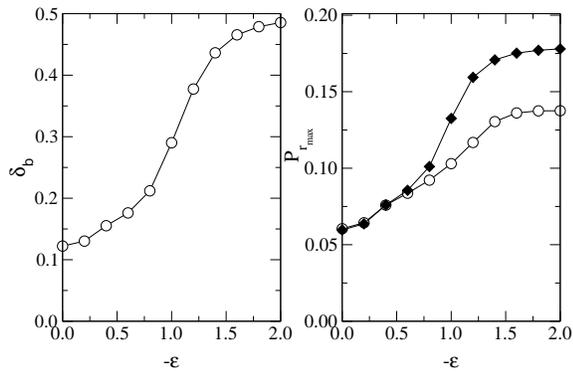}
\caption{Results for the $4\times 6$ cluster, $n=0.25$,
$U=V=0$, $t_b=t_f=1$, $\lambda_{A,s}=0.33$, $\lambda_{A,h}=0.06$.
(a) Boson density, and (b) pair-pair correlations at the maximum
distance along a bosonic chain (filled triangles) and perpendicular
to the chains (open circles)
versus the absolute value bosonic site energy.
}
\label{figstrip}
\end{center}
\end{figure}

\subsection{A ``toy" model of stripes}

With respect to the application of the present model to the
stripe phase in the cuprates, one should take into account that
the $t$-$J$ model considered as a microscopic model from
which the couplings in Fig.~\ref{fig1}(d) were derived, was
meant to be the strong-coupling limit of the one-band Hubbard
model. In these models pairs involve {\em electrons} with opposite
spins. In the $t$-$J$ model, as applied to cuprates, pairing
involve doped holes which are described by 
singlets.\cite{zhangrice} Although a derivation of an effective
model from this version of the $t$-$J$ model is possible, it is
instructive to use the already obtained results to study
a ``toy" model of stripes. To do this, a fermion should have
the meaning of a doped hole, and the half-filled state of the
cuprates should be the ``vacuum" of model (\ref{hambf}). From
Fig.~\ref{fig1}(d), the approximate values of the couplings
are $t_b=t_f=1$, $\lambda_{A,s}=0.33$, $\lambda_{A,h}=0.06$,
$U=V=0$. On the $4\times 6$ cluster described above, in order that
the stripes be at a linear filling of one quarter, there should be
6 doped holes, corresponding to a doping on the original cluster
of 0.125. In this study the variable is the site energy at the
stripes, $\epsilon$, which may also depend of various
mechanisms such as structural details, phonons.\cite{rierastrip}
Results are shown in Fig.~\ref{figstrip}. The main conclusion
is that pairing is enhanced in both longitudinal and transversal
directions, even though doped holes are increasingly localized at
the ``stripes".

\section{conclusions}
\label{conclusions}

In summary, a model of coupled bosonic and fermionic
chains was proposed to describe the physics of compounds in which
pairing takes place in quasi-1D structures such as two-leg ladders.
Starting from a microscopic model, an exact projection procedure
on a small cluster suggests that a pair splitting kind of Andreev
process is related to the physics of repulsive $U$ systems,
characterized by d-wave pairing, while a pair hopping Andreev
process is more related to a negative $U$ kind of physics leading
to s-wave pairing. This elementary projection also gives 
indication of how the effective couplings are changed with 
pressure. The values of these couplings for a specific compound
should be obtained from a realistic, in general complex, microscopic
model, and in this case an exact diagonalization procedure dealing
with much larger clusters than the ones of Fig.~\ref{fig1}(c) should
be used. Although in principle both types of Andreev couplings are
going to be present in the effective model irrespective of the
nature of the pairing, the purpose of the present study
was to determine the more general and important properties of those
processes taken {\em separately}. The conclusion was that,
close to quarter-filling, a pair
splitting process is more efficient to enhance long-distance
pairing than pair hopping from the superconducting to the non
superconducting chains. So far this result would suggest that if
the effect of pressure translates in an increasing of 
$\lambda_A$ in, for example, $\beta$-Na$_{0.33}$V$_2$O$_5$, then
the presence of pairing would be more likely
the result of strongly correlated electron physics in these
compounds. More detailed predictions would require to determine
the effective couplings more precisely as discussed above.
In any case, even at this general level, more predictions
for example for ARPES experiments could be obtained by
studying dynamical properties.\cite{rierafut}

The use of computers at the 
Institute for Materials Research (IMR), Tohoku University, Japan,
is gratefully acknowledged.


\begin{thebibliography}{}
\bibitem{note1}The symbol $\lambda_A$ is used to denote
    indistinctly $\lambda_{A,s}$ or $\lambda_{A,h}$.
\end{thebibliography}
\end{document}